# Field Fractal Cosmological Model
# As an Example of Practical Cosmology Approach

© Yu. Baryshev[1,2]

[1] Astronomical Institute of the St.-Petersburg State University, St.-Petersburg, Russia
[2] Email: yubaryshev@mail.ru

**Abstract:**
The idea of the global gravitational effect as the source of cosmological redshift was considered by de Sitter (1916, 1917), Eddington (1923), Tolman (1929) and Bondi (1947). Also Hubble (1929) called the discovered distance-redshift relation as "De Sitter effect". For homogeneous matter distribution cosmological gravitational redshift is proportional to square of distance: $z\_grav \sim r^2$. However for a fractal matter distribution having the fractal dimension D=2 the global gravitational redshift is the linear function of distance: $z\_grav \sim r$, which gives possibility for interpretation of the Hubble law without the space expansion. Here the field gravity fractal cosmological model (FGF) is presented, which based on two initial principles. The first assumption is that the Feynman's field gravity approach describes the gravitational interaction, which delivers a natural basis for the conceptual unity of all fundamental physical interactions within the framework of the relativistic and quantum fields in Minkowski space. The second hypothesis is that the spatial distribution of gravitating matter is a fractal at all scales up to the Hubble radius. The fractal dimension of matter distribution is assumed to be D = 2, which implies that the global gravitational redshift is the explanation of the observed linear Hubble law. In the frame of the FGF all three phenomena - the cosmic background radiation, the fractal large scale structure, and the Hubble law, - could be the consequence of a unique large scale structure evolution process of the initially homogeneous ordinary matter without nonbaryonic matter and dark energy.

## 1. Initial hypotheses of the Field Gravity Fractal cosmology framework

The basic feature of the modern Standard Cosmological Model (SCM) is that the observed luminous matter (galaxies), which is strongly clustered in the Universe, contributes only 0.5 percent of the total homogeneously distributed "unseen matter". Hence in the general relativistic homogeneous world model (Friedmann model) the uniform nonbaryonic dark matter and dark energy determine the dynamics of the whole Universe, while the ordinary baryonic matter (which is actually observed) is dynamically unimportant in the global evolution of the Universe.

Another obstacle of the SCM is that the space expansion paradigm leads to the deep conceptual problems of the cosmological physics, such as continuous vacuum creation, violation of energy conservation within each comoving volume, non-Doppler character of the expanding space cosmological redshift (Baryshev 2008a).

Here the field gravity fractal cosmological model (FGF) is presented, which is free from these SCM conundrums. FGF is based on the two natural ideas – the first is that the actual baryonic matter distribution is well described by a fractal density law and the second is that the gravitational interaction is described by the Feynman's field approach to the gravity theory (Feynman et al. 1995), which has passed all weak-field tests and will be tested soon in strong-gravity regimes (Baryshev 2008b,c). Therefore it is natural to inspect as an alternative cosmological framework the field-gravity fractal cosmological model, though the FGFM is still a developing subject. Within its framework a new qualitative picture of the Universe has emerged, with some quantitative results that may be tested by current and forthcoming observations.

*The first assumption* is that the Feynman's field gravity approach is used for description of the gravitational interaction. The strategy and basic principles of the field approach to gravitation were discussed by Feynman, who emphasized that for gravity theory "geometrical interpretation is not really necessary or essential for physics" (Feynman, Morinigo & Wagner 1995, p. 113).The field approach delivers a natural basis for the conceptual unity of all fundamental physical interactions, within the framework of the relativistic and quantum fields in Minkowski space. It also gives a possibility to consider matter distributions in the infinite non-expanding Minkowski space without gravitational potential paradox.

*The second hypothesis* is the fractal distribution of dark matter from the scales of galactic halos up to the Hubble radius. The fractal dimension of the total (luminous and dark) matter distribution is assumed to be D = 2 , and the global gravitational redshift explains the observed linear Hubble law.

It is interesting to derive predictions for the classical tests in this model whose basis differs radically from the standard model in two ways: gravitation is described by the field instead of geometry, and matter composition and distribution is a fractal baryonic dark matter instead of homogeneous dark energy.

## 2. Cosmological solution in the field gravity

Usually field gravity is treated only for weak-field approximation using the iteration procedure (Baryshev 2008b,c). A specific feature of the field gravity theory is that there is the case of a weak force with $\nabla \varphi \to 0$ while $|\varphi| \to c^2/2$. This is what happens in the cosmological problem and we can obtain some quantitative results even at the post-Newtonian level.

Let us consider the case of a static homogeneous ($\rho$ = const) dust-like cold (p=0, e=0) matter distribution within infinite space. Using expressions for the post-Newtonian EMTs of matter and taking into account the traceless of the field and interaction EMTs, we get the equation for the $\psi^{00} \equiv \varphi$ component in the form

$$\Delta \varphi = 4\pi G\, [\rho + \frac{2}{c^2}\rho\varphi + \frac{2}{8\pi G c^2}(\vec{\nabla}\varphi)^2]\ . \tag{1}$$

In our case the main terms in the right-hand side of Eq.(1) are the positive rest mass density $\rho$ and the negative interaction mass density $(2\rho\varphi)/c^2$. The last term can be neglected, because for $\varphi \to const$ its gradient $\nabla\varphi \to 0$. Hence we have the simple equation

$$\Delta \varphi - \frac{8\pi G \rho}{c^2} \varphi = 4\pi G\rho \quad \text{(which is equivalent to Einstein's } \Lambda\text{-equation)} \quad \Delta\varphi - \Lambda\varphi = 4\pi G\rho\ , \tag{2}$$

and we may conclude that the $\Lambda$-term in the field gravity theory is

$$\Lambda = \frac{8\pi G \rho}{c^2}\ , \tag{3}$$

and it comes from the contribution of the energy-momentum tensor of the interaction. Therefore the cosmological solution of Eq.(2) with (3) is

$$\varphi = -\frac{4\pi G\rho}{\Lambda} = -\frac{c^2}{2}\ . \tag{4}$$

So within the field gravity there is the unique natural static cosmological solution for the case of global homogeneous distribution of matter in the infinite space.

## 3. Fractal matter ball with finite radius

The cosmological solution (4) can be also derived as a limiting case ($r \to \infty$) of the exact solution of Eq.(2) for a matter ball with radius $r$.

### *Fractal dimension D = 3*

For a homogeneous matter distribution $\rho$ = const the solution of Eq.(2) inside the ball has the form (Baryshev & Kovalevskij 1990):

$$\frac{\varphi(x)}{c^2} = -\frac{1}{2} + \frac{sh(x)}{2x\, ch(x_0)}\ . \tag{5}$$

Here $x = r/R_H$ is the dimensionless radius of the ball in units of the Hubble radius $R_H = c/t_H = c/(8\pi G\rho)^{1/2}$, and $x_0 = r_0/R_H$ where $r_0$ is the radius of the ball. The gravitating mass of this ball is

$$M(x) = M_H\, x\, (1 - \frac{th(x)}{x})\ ,$$

where $M_H = R_H c^2 / 2G$ is the Hubble mass, which is the characteristic mass within the Hubble radius.

For sufficiently small distances ($r << R_H$), the gravitational potential has Newtonian behavior, and for large distances ($r >> R_H$) the mass grows linearly so that the gravitational potential in the center of the ball asymptotically reaches the value ($-c^2/2$).

The constant gravitational potential in the cosmological solution resolves the long standing paradox of Jeans on cosmological initial values. Now $\varphi = const$ is consistent with an infinite initially homogeneous gas distribution.

*Fractal dimension D = 2*

In the case of the fractal dark matter distribution with D = 2 the rest mass density law is $\rho(r) = \rho_0 r_0 / r$ and the solution of Eq.(2) inside the ball has the form (Nagirner 2006):

$$\frac{\varphi(x)}{c^2} = -\frac{1}{2} + \frac{1}{\sqrt{x}} \left[ C_1 I_1(4\sqrt{x}) + C_2 K_1(4\sqrt{x}) \right] , \tag{6}$$

where $I_1$, $K_1$ are the modified Bessel functions and $x$ is the dimensionless distance. Using ordinary boundary conditions for the gravitational potential of a finite ball with radius $x = x_0$ one finds that $C_2 = 0$ and $C_1 = 1/(4 I_0(4\sqrt{x_0}))$, where $I_0(x)$ is the modified Bessel function.

The total gravitating mass inside the fractal ball of radius $r$ is:

$$M(x) = M_H \, x \, [1 - \frac{I_1(4\sqrt{x})}{2\sqrt{x} \, I_0(4\sqrt{x})}] . \tag{7}$$

Here $x = r/R_H$ is the dimensionless radius in units of the Hubble radius $R_H = c^2/(2\pi G \rho_0 r_0)$, and the product $\rho_0 r_0$ is a new fundamental constant which is defined by the lower cutoffs $\rho_0$ and $r_0$ of the fractal structure with D = 2, $M_H = R_H c^2 / 2G$ is the Hubble mass as above.

**4. Cosmological gravitational redshift**

In static space, filled by infinitely distributed matter, the cosmological redshift may appear as a global gravitational effect due to the mass of the ball centered at the light source with radius equals to the distance between the source and the observer.

*De Sitter effect of gravitational redshift*

In early history of the relativistic cosmology de Sitter (1916, 1917), Eddington (1923) and Tolman (1929) discussed the possibility to observe the *de Sitter effect* in a static cosmological model. De Sitter (1917) found a static solution of Einstein's equations for an empty universe with cosmological constant $\Lambda$:

$$ds^2 = (1 - \frac{r^2}{R_\Lambda^2}) c^2 dt^2 - \frac{dr^2}{1 - \frac{r^2}{R_\Lambda^2}} - r^2 (d\theta^2 + \sin^2\theta \, d\phi^2) . \tag{8}$$

Here $r$ is the distance from the source to the observer, and $R_\Lambda^2 = 3/\Lambda$ is the characteristic radius corresponding to the cosmological constant $\Lambda$. The de Sitter effect is caused by the $g_{00}$ component of the metric and according to the definition ($1 + z_g = 1/\sqrt{g_{00}}$) is the cosmological gravitational redshift for a homogeneously distributed substance with positive mass density $\rho_\Lambda = \Lambda c^2 /(8\pi G)$. Notably, Einstein (1917) constructed his first cosmological model with the ad hoc extra condition $g_{00} = 1$. Thus he lost the gravitational redshift in his static model, which was later rediscovered by Bondi (1947).

Eddington (1923) emphasized that in ``De Sitter's theory... there is the general displacement of spectral lines to the red in distant objects due to the slowing down of atomic vibrations which... would be erroneously interpreted as a motion of recession''. In fact, in his famous study Hubble (1929) refers to the de Sitter effect as an explanation of the discovered cosmological redshift effect and the distance-redshift law (Sandage 1975; Smith 1979).

In fact this is a new physical effect (de Sitter-Bondi effect), which appears in cosmology, due to the non-locality of cosmological observations. It appears only on cosmological scales and is not the Pound-Rebka experiment probing the local gravity field. So the cosmological gravitational redshift was considered as an explanation of observed spectral shifts already before the space expansion interpretation.

The cosmological gravitational effect differs from the local gravitational effect, which occur when there is the preferred center of gravity and a photon may have both possibilities - redshift or blueshift, depending on the direction of propagation (from the center or to the center). However in cosmological case all points "alike" and spectral shift is always redshift because the center of the cosmological ball is always in the emitter of the photon, and an observer at the surface (causality principle).

*Small redshifts*

Within expanding space cosmology Bondi (1947) rediscovered the de Sitter effect, when he demonstrated that for a homogeneous matter distribution and small redshifts ($z << 1$) the gravitational cosmological redshift is:

$$z_{\cos-grav} = \frac{\delta\varphi(r)}{c^2} = \frac{1}{2}\frac{GM(r)}{c^2 r} = \frac{1}{4}\Omega_0 \left(\frac{r}{R_H}\right)^2 , \tag{9}$$

where $\delta\varphi = \varphi(r) - \varphi(0)$ is the gravitational potential difference between the surface and the center of the ball, and $R_H = c/H_0$ is the Hubble distance.

Why does the cosmological gravitational effect give the redshift? From the causality principle it follows that the event of emission of a photon (or a spherical wave) by the source, which marks the centre of the ball, must precede the event of detection of the photon by an observer. The latter event marks the spherical edge where all potential observers are situated after the transition time $t = r/c$. Therefore to calculate the cosmological gravitational shift within the cosmologically distributed matter one should cut a material ball with the center in the source and with the radius of the ball equal to the distance between the source and an observer. In this case the cosmological gravitational shift is towards red.

It is true that in some discussions the observer was put to the center of the ball and hence a blueshift was obtained instead of Bondi's and de Sitter's redshift ( Zeldovich & Novikov 1984, p.97; Peacock 1999, problem 3.4). However, such a choice of the reference frame violates the causality in the process considered: the ball with the source on its surface has no causal relation to the emission of the photon.

Note that from Eq.(9) we see that when $c \to \infty$, the redshift drops to zero. Indeed, in Newtonian physics one may choose the sphere either around the source or the observer, without causality problems, and thus infer that $z_{\cos-grav} = -z_{\cos-grav} = 0$. Hence the global gravitational redshift is essentially relativistic effect.

*Fractal matter distribution*

Within the homogeneous matter distribution the global gravitational redshift is proportional to square of the distance between the source and observer ( $z \propto r^2$ ). In order to have a linear redshift-distance relation within universe with no preferred center one may consider a fractal distribution with fractal dimension D=2 (Baryshev 1981). Indeed, for a fractal distribution where $M(r) \propto r^D$ one may derive for z<<1 the following relation for the gravitational part of the cosmological redshift within the fractal galaxy distribution:

$$z_{\cos-grav} = \frac{4\pi G\rho_0 r_0^2}{c^2 D(D-1)} (\frac{r}{r_0})^{D-1} . \tag{10}$$

Here $\rho_0$ and $r_0$ are the density and radius of the zero level of the fractal structure. For the case D = 2 the density is $\rho(r) = \rho_0 r_0 / r$ and the mass of the ball is $M(r) = 2\pi \rho_0 r_0 r^2$, hence the cosmological gravitational redshift is a linear function of distance:

$$z_{\cos-grav} = \frac{2\pi G\rho_0 r_0}{c^2} r = \frac{H_g}{c} r . \tag{11}$$

*The gravitational Hubble constant $H_g$ may be expressed as*

$$H_g = 2\pi \rho_0 r_0 \frac{G}{c} \quad . \tag{12}$$

For a structure with fractal dimension D = 2 the constant $\beta = \rho_0 r_0$ may be actually viewed as a new fundamental physical constant which determining the value of the gravitational Hubble constant. If the value of the fractal constant is $\beta = 1/(2\pi) \ (g/cm^2)$, e.g. $\rho_0 = 5.2 \times 10^{-24} \ (g/cm^3)$, and $r_0 = 10 \ kpc$, then $H_g = 2\pi \beta G/c = 68.7 \ (km/s)/Mpc$. Intriguingly the fractal law $\rho(r) = \rho_0 r_0 / r$ and possible gravitational Hubble constant $H_g$ was derived by Baryshev&Raikov (1988) using Large Numbers approach, where $\rho_0$ and $r_0$ can be expressed via the fundamental microphysical constants *G, h, c, m_p*. So the universal linear gravitational redshift law within the fractal structure with D=2 would have deep roots in the fundamental physics. It also require the construction of the future *G-h-c* gravitation theory.

*Large redshifts*

For the case of large redshifts there is, unfortunately, still no exact field gravity theory and we consider only some hypothetical approximate formulas. From the PN approximation we may surmise that the strong gravity redshift is given by the relation $1 + z = 1/\sqrt{1 + 2\varphi(r)/c^2}$, which describes a spectral line shift for an atom radiating a photon at point *r*, which is detected by an observer at infinity. Hence for the cosmological case of a source at the center of a matter ball (*r = 0*) and an observer at the surface of the ball (*r = R*), the observed redshift will be

$$1 + z_{obs} = \frac{1 + z(0)}{1 + z(R)} = \left(\frac{1 + 2\varphi(R)/c^2}{1 + 2\varphi(0)/c^2}\right)^{1/2} \quad . \tag{13}$$

Inserting the expressions for the gravitational potential one can derive the following formula for the cosmological gravitational redshift -- distance relation:

$$z_{obs}(x) = \left[\frac{1}{2\sqrt{x}} I_1(4\sqrt{x})\right]^{1/2} - 1 \equiv W(x) \quad . \tag{14}$$

Here $x = r/R_H$, $R_H = c/H_g$, $I_1(y)$ is the modified Bessel function.

### 5. Specific features of the fractal framework

*The total mass-radius relation*

Eq.(7) for the gravitating mass has two characteristic limiting cases. For small distances ( $r \ll R_H$, $x \ll 1$ )

$$M(r) = 2\pi \rho_0 r_0 \, r^2 = 4.8 \times 10^{11} \, M_{Sun} (\frac{r}{10 \ kpc})^2 \quad . \tag{15}$$

Note the interesting coincidence that this mass is close to a total galaxy mass (including dark matter) within the radius *r* about 10 kpc, and also to the mass of the galaxy Universe within the Hubble radius $r \approx R_H$, having the mass density close to the critical value.

Therefore to produce the gravitational Hubble law on scales of about 10 Mpc the total mass within such balls should be $M(10 Mpc) = 4.8 \times 10^{16} \, M_{Sun}$, and within 1 Gpc $M(1 Gpc) = 4.8 \times 10^{21} \, M_{Sun}$. Such values much exceed the mass of the luminous matter and this is why the FGF model is compelled to assume that a sufficient amount of dark matter has the fractal distribution with D = 2. To have sufficiently small fluctuations in the Hubble law in different directions around an observer the fractal should be a special class: isotropic with small lacunarity.

The observed distribution of luminous matter (galaxies) on scales from 10 kpc up to 100 Mpc may be approximated by a fractal distribution with D = 2. This means that within the FGF model both dark and lu-minous matter is similarly distributed on these scales. The nature of the fractal dark matter has to be determined from future observations. Current restrictions on possible dark matter candidates leave room for cold dead stars, neutron and quark stars, Jupiters, planet size objects, asteroids and comets, Pfeniger's hydrogen cloudlets, and also quark dust (V.V.Sokolov's suggestion), i.e. quark bags with small masses. Also different kinds of cold non-baryonic dark matter might make fractal dark matter structures.

For large distances ($r \gg R_H$, $x \gg 1$) the total relativistic mass is

$$M(r) = \frac{c^2}{2G} r \qquad (16)$$

for both D = 3 and D = 2 fractal structures. This means that at large distances both distributions produce similar gravitating mass. Because the gravity force goes to zero on scales larger than $R_H$, the fractal dimension of dark matter may become D=3, corresponding to a homogeneous distribution.

It is critical for the fractal framework that the existence of the global gravitational redshift should be supported both theoretically and observationally, as well as its ability to produce the observed linear distance--redshift law. While crucial studies around these fundamental questions go on, one should also address other cosmological key questions: structure formation and the background radiation. Are there prospects to understand these outside the traditional scope of the big bang model? We think ``yes'', following from tentative considerations.

### *The evolution of the Universe*

In Minkowski space-time filled by matter there is a special frame of reference, namely the one where the matter is at rest on the average relative to the cosmic background radiation. This frame of reference allows one to speak also about a universal time and the arrow of time is determined by the growth of the local entropy. Initial fluctuations in the homogeneous gas of primordial hydrogen exponentially grow into large-scale structures according to the classical scenarios by Jeans (1929) and Hoyle (1953).

The fractal structure of matter distribution with D = 2 could naturally originate as the result of the evolution of the initial fluctuations within the explosion scenario (Schulman & Seiden 1986). The fractal dimension D = 2 is also preferred in the dynamical evolution of self-gravitating N-body system (Perdang 1990; de Vega et al., 1996; 1998). Recently new kinds of arguments for the privileged value D = 2 were presented by Mureika (2006). In a geometric re-interpretation of large-scale structure he introduced the concept of fractal holography, related to current theories of holographic cosmologies.

Within a D = 2 fractal structure the gravity force acting on a particle is constant because of $M \propto r^2 \Rightarrow F \propto M/r^2 \propto const$. The positive energy density of the gravity field within D = 2 fractal structure is also constant: $\varepsilon_g = T_g^{00} \propto (d\varphi/dr)^2 \propto const$. This corresponds to a homogeneous distribution of the "gas" of virtual gravitons. Three is an interesting suggestion by Raikov (2008) that the Pioneer's effect in solar system may be caused by these cosmological gravitons.

The time-scale of the structure evolution is determined by the characteristic Hubble time:

$$t_H \approx R_H / c \propto 1/\sqrt{\rho_H} \approx 10^{10} \; yrs \qquad (17)$$

The total evolution time of the Universe may be several orders of magnitude larger, which could be tested by observations at high redshifts and by numerical simulations of the large-scale structure and galaxy formation in static space but dynamically evolving matter.

### *The cosmic microwave background radiation*

According to the classical argument by Hoyle (1982, 1991) the CBR could be a remnant of the evolution of stars, which energy density equals to the energy released by the nuclear reactions in stars of all generations. The optical photons radiated by stars could be thermalized by scattering and gravitational deflections by structures of different masses and scales.

The fractal dark matter is also a product of the process of stellar evolution and large scale structure formation. Hence in the frame of the FGF all three phenomena - the cosmic background radiation, the fractal large scale structure, and the Hubble law, - could be consequences of a unique evolution process of the initially homogeneous cold hydrogen gas.

## 6. Main cosmological parameters and relations

The possible gravitational cosmological redshift in Minkowski space prompts one to reanalyze the relations between proper metric, angular, and luminosity distances.

### Source of radiation at distance r

There are two main reference frames for description of the cosmological observations: the observer's local inertial frame and the source's local inertial frame. Due to the universality of physical laws in the Universe (assumed as a principle) all local processes are identical, e.g., the hydrogen atoms are everywhere the same as on the Earth.

The non-locality of cosmological observation originates when one compares a photon from a distant source, radiated in its local system, with the photon of the local observer, radiated in his local system. Here profound new cosmological physics enters the scene and we present below a few examples of its operation. The non-locality leads to fundamental apparent changes in the measured source parameters.

Let us consider a spherical source of radiation in its local inertial frame, where the units of length "cm", time "sec" and mass "gram" are defined, so one can measure the intensity of radiation "ergs/(cm^2Hz ster sec), the bolometric luminosity "ergs/sec", and the linear sizes of objects in "cm". At the proper metric distance $r$ there is an observer who measures the redshift, the angular size, the flux and the surface brightness of the source.

### General distance - redshift relation

From Eq.(14) for the cosmological gravitational redshift one can write the general distance-redshift relation in the form

$$r_{metr}(z) = R_H Y_g(z) \quad \text{with} \quad Y_g(z) \equiv W^{-1}(x) \tag{18}$$

where $r$ is the metric distance to an object having the redshift $z$, Y(z) is the inverse function of W(x) defined by (14). So for z << 1 the Eq.(18) gives the linear Hubble law $r \approx R_H z$, and for z >> 1 the distance is proportional to $(\ln z)^2$.

In the FGF model the redshift of a distant source is a measure of global mass cosmologically distributed in the sphere around the source. This "de Sitter-like effect" may be understood as a consequence of non-locality of the cosmological observations, producing an apparent slow-down of all local processes seen by a distant observer. Note that observers at different distances from the same source will see different redshifts of this source, while actual local processes are the same in all local systems. This may point to the principle of relativity of the cosmological gravitational potential, introduced by Einstein (1911) in his early study of gravity.

### The angular size - redshift relation

The angular size $\theta$ of a source with linear size $d$ at the metric distance $r$ is defined by the Euclidean relation $\theta = d/r_{ang}$ which gives definition of the angular distance. There are several possibilities to introduce the relation between metric and angular distances and we unify notations by introducing a parameter $n$ as an exponent in the formula:

$$r_{ang} = \frac{r_{metr}}{(1+z)^n}, \tag{19}$$

where $n$ may be 0 or 1 depending on chosen physical possibility. Hence the angular size-redshift relation has the form

$$\theta(z) = \frac{d}{R_H} \frac{(1+z)^n}{Y_g(z)}. \tag{20}$$

### Apparent luminosity and surface brightness

The bolometric luminosity distance $r_{lum}$ is defined by the relation for the observed flux from a source $F_{obs} = L/4\pi r_{lum}^2$. There are several physical possibilities for definition of the luminosity distance, which we describe by the exponent $m$ in the relation:

$$r_{lum} = r_{metr}(1+z)^m, \tag{21}$$

where $m$ can be equal to 1 or 2.

Taking into account that $m_{bol} = -2.5 \lg F_{bol} + const$, the bolometric apparent magnitude – redshift relation will be

$$m_{bol}(z) = 5\lg(Y_g(z)(1+z)^m) + const. \tag{22}$$

The surface brightness $J$ as the ratio of the observed flux to the square of the angular size will be:

$$J(z) = \frac{F(z)}{\theta^2(z)} = \frac{J_0}{(1+z)^{2(m+n)}}, \tag{23}$$

The general relation between metric, angular and luminosity distances may be written in the form

$$r_{metr} = r_{ang}(1+z)^n = \frac{r_{lum}}{(1+z)^m}. \tag{24}$$

For example if m = n =1, we get surface brightness is $J(z) = J_0/(1+z)^4$.

## 5. Conclusion: crucial tests of field gravity fractal cosmological model

Note that the FGF framework is a developing subject and still contains many open questions. In its present preliminary state it may be considered as an example for Practical Cosmology in constructing possible cosmological physics, which is free from paradoxes of the expanding space of the contemporary standard cosmological model (Baryshev 2008a).

The role of crucial tests for the field gravity fractal cosmological model will play the laboratory and astrophysical tests of the gravity physics. The most promising observational cosmological test for distinction between SCM and FGF is the direct observational proof or disproof the reality of space expansion, like Sandage's (1962) suggestion for measuring the change of redshift with time which is one of the main goals of the OWL ESO telescope (Pasquini et al. 2005).